\documentclass[reprint]{revtex4-2}

\usepackage{graphicx}
\usepackage{dcolumn}
\usepackage{bm}
\usepackage{braket}
\usepackage{amsmath}
\usepackage{amssymb}
\usepackage{float}

\begin{document}

\preprint{APS/123-QED}

\title{Non-Markovian stochastic Gross-Pitaevskii equation for the exciton-polariton Bose-Einstein condensate}

\author{A. D. Alliluev}
\affiliation{V. I. Il'ichev Pacific Oceanological Institute, Far East Branch of the Russian Academy of Sciences, Vladivostok 690041, Russia}
\author{D.V. Makarov}
\affiliation{V. I. Il'ichev Pacific Oceanological Institute, Far East Branch of the Russian Academy of Sciences, Vladivostok 690041, Russia}
\author{N.A. Asriyan}
\affiliation{N.L. Dukhov Research Institute of Automatics (VNIIA), Moscow 127030, Russia}
\author{A.A. Elistratov}
\affiliation{N.L. Dukhov Research Institute of Automatics (VNIIA), Moscow 127030, Russia}
\author{Yu. E. Lozovik}
\affiliation{Institute for Spectroscopy RAS, Troitsk 108840, Moscow, Russia}
\affiliation{MIEM, National Research University Higher School of Economics, Moscow 101000, Russia}

\begin{abstract}
In this paper, a non-Markovian Gross-Pitaevskii equation is proposed to describe the formation of a condensate in an exciton-polariton system under incoherent pumping.
By introducing spatially delta-correlated noise terms, we observe a transition from a spatially ordered phase to a disordered one as the temperature increases.
In course of this process, the population of the condensate is significantly reduced.
Irregularly located separate dense spots of condensate above the transition temperature are revealed. Using the Gabor transform, it is shown that, with increasing temperature, the condensate decoheres, that is accompanied by the transition from narrowband to broadband spectral density.
 
\end{abstract}

\maketitle


\section{Introduction}
Open quantum systems attract much interest in recent years, which is largely due to the rapid development of quantum computing technologies. A common approach for studying such systems is the Markov approximation \textit{i.e.} using the Lindblad equation for the density matrix
or the stochastic Schr\"odinger equation \cite{Petruccione_Breuer,Kol_Shep}. 

When dealing with a condensate of quasiparticles, we can consider it as an open quantum system linked to a reservoir of noncondensed particles. For a theoretical description, the stochastic Gross-Pitaevskii equation can be used \cite{Stoof,Cockburn_Proukakis}. However, due to the significant development of experimental methods, the reservoir may have a relatively low temperature as well as a narrow energy spectrum. This makes Markovian approximation invalid for such systems. Non-Markovian evolution infers existence of memory effects in the system. The corresponding evolution equation is an integro-differential one with time non-local terms \cite{DeVega}. For instance, Nakajima-Zwanzig \cite{Nakajima,Zwanzig} equation or non-Markovian versions of the Lindblad equation \cite{Breuer04,Breuer-RMP} are used for the density matrix. Though the non-Markovian Schrödinger equation for non-interacting particles was first introduced in \cite{Diosi_Strunz}, its generalisation for systems with interparticle interaction is still a challenge. Significant progress has been achieved in \cite{EL18}, where, using the Keldysh diagram technique, a system of non-Markovian equations for an exciton-polariton Bose-Einstein condensate in a Fabry-Perot optical microresonator has been derived.

An exciton-polariton is a bound state of an exciton and a photon, the lifetime of which is significantly affected by the quality factor of the microresonator. Exciton-polariton condensation arises as a result of their thermalization stimulated by laser pumping. Exciton polaritons are attractive for technological applications largely because of the prospects for their use in quantum computing applications \cite{Kavokin}, as well as in the context of the polariton laser \cite{Kim}.

In \cite{EL18} it was assumed that the condensate corresponds to a macroscopically populated state with zero momentum.
Therefore, the resulting system of equations can be treated a discrete non-Markovian Gross-Pitaevskii equation. Unfortunately, a rigorous derivation of a similar Gross-Pitaevskii equation with spatial dynamics taken into account remains challenging. However, as a reasonable approximation, we can use the evolution equation from \cite{EL18} a small neighborhood of zero momentum. It corresponds to a smoothly varying spatial profile of the macroscopic wave function of the condensate. Under this assumption one can propose a non-Markovian version of the Gross-Pitaevskii equation for the exciton-polariton condensate.

The paper is organized as follows. In the next section we introduce the non-Markovian stochastic Gross-Pitaevskii equation. The process of condensate formation is studied in Section \ref{sec:Numer} by means of numerical simulation. 
In Conclusion we summarize and discuss the results obtained.

\section{Evolution equations}
\label{sec:Theory}

When dealing with exciton-polariton condensation one may usually neglect the population of the upper spectral branch (upper polariton states) \cite{Gavrilov}. It leads to the following form of the Gross-Pitaevskii equation:
\begin{equation}
\begin{aligned}
 i\hbar\frac{\partial\psi(\mathbf{r},t)}{\partial t} =
 \hat H_0\psi(\mathbf{r},t) + P_{\mathrm{coh}}(\mathbf{r},t) + \\
 \hat D_{\text{cav}}\psi(\mathbf{r},t)
 + \hat D_{\text{ex}}\psi(\mathbf{r},t),
\end{aligned}
\label{sys0}
 \end{equation}
where $\hat H_0$ is an unitary operator including kinetic and potential energies, and energy corrections due to interparticle interactions,
\begin{equation}
 \hat H_0 = -\frac{\hbar^2}{2m^*}\nabla^2 + V(\mathbf{r}) 
 + \alpha_{\mathrm{c}}|\psi(\mathbf{r},t)|^2 + \alpha_{\mathrm{r}}\rho_{\text{r}}(\mathbf{r},t).
\end{equation}
Here $m^*$ stands for the effective polariton mass, which is approximately $10^{-5}$--$10^{-4}$ $m_{\text{e}}$, $V(\mathbf{r})$ is the external potential, $\alpha_{\text{c}}$ is the interaction constant for condensate polaritons, $\alpha_{\text{r}}$ describes condensate-reservoir interaction, $\rho_{\text{r}}(\mathbf{r},t)$ is the exciton reservoir density. The function $P_{\mathrm{coh}}(\mathbf{r},t)$ in 
(\ref{sys0}) describes coherent laser pumping of microresonator photons.

Non-Hermitian operators $\hat D_{\text{cav}}$ and $\hat D_{\text{ex}}$ correspond to 
interactions with the photon and exciton reservoirs respectively. Hereafter we assume that the interaction with the photon reservoir is Markovian. This assumption leads to
\begin{equation}
 \hat D_{\mathrm{cav}}\psi = -i\gamma_{\mathrm{cav}}\psi + \eta_{\mathrm{cav}}(\mathbf{r},t).
\end{equation}
Using the truncated Wigner approximation, one can derive the following expression for the autocorrelation function:
\begin{equation}
 \braket{\eta^*_{\mathrm{cav}}(\mathbf{r},t)\eta_{\mathrm{cav}}(\mathbf{r'},t')} = 
 \frac{\gamma_{\mathrm{cav}}}{\Delta x\Delta y}\delta(\mathbf{r}-\mathbf{r'})\delta(t-t'),
\end{equation}
with $\Delta x$ and $\Delta y$ being the grid cell sizes.

The non-Markovian behaviour of the system is mainly associated with the interaction with the exciton reservoir. Considering only the low-momentum states as a condensate, we can use the following approximation
\begin{equation}
 \hat D_{\text{ex}}\psi(\mathbf{r}) \simeq \hbar\int\limits_{0}^{t}dt' \Sigma^{\text{R}}(t,t')\psi(\mathbf{r},t')	
 + \eta_{\text{ex}}(\mathbf{r},t),
\end{equation}
\textit{i.e.} treat the retarded self-energy term $\Sigma^{\text{R}}(t,t')$ as a spatially homogeneous one. We use expressions derived in \cite{EL18} 
\begin{equation}	
  \Sigma^{\text{R}}(t,t') = i\frac{\rho_{\text{r}}^2\alpha_{\text{c}}^2 }{\hbar^2}\frac{e^{-\gamma_{\text{ex}} (t-t')}}{1 + \left[\frac{k_{\text{B}}T}{\hbar}(t-t')\right]^2}\theta(t-t'),
  \label{retarded}
\end{equation}
with $\theta(t)$ being the Heaviside function, $\gamma_{\text{ex}}$ -- condensate exciton decay rate. The corresponding expression for the Keldysh component of the self-energy term is as follows:
\begin{equation}	
  \Sigma^{\text{K}}(t,t') = 
  -i\frac{\rho_{\text{r}}^2\alpha_{\text{c}}^2 }{\hbar^2}\frac{e^{-\gamma_{\text{ex}} |t-t'|}}{1 + \left[\frac{k_{\text{B}}T}{\hbar}(t-t')\right]^2}.
  \label{keldysh}
\end{equation}
This expression defines the temporal autocorrelation function of exciton fluctuations:
\begin{equation}
 \braket{\eta^*(\mathbf{r}, t)\eta(\mathbf{r'},t')} = i\hbar^2\delta(\mathbf{r},\mathbf{r'})\Sigma^{\text{K}}(t,t').
 \label{autocorr}
\end{equation}

When modeling the spatial structure of exciton fluctuations, we utilize above a simple approximation, considering them as delta-correlated. Condensate formation with correlated noise was studied in \cite{PLA22}.
Thus, equation (\ref{sys0}) takes the form
\begin{align}
 i\hbar\frac{\partial\psi(\mathbf{r},t)}{\partial t} =
 \hat H_0\psi(\mathbf{r},t) - \frac{i\hbar \gamma_{\text{cav}}}{2}\psi(\mathbf{r},t) + P_{\mathrm{coh}}(\mathbf{r},t)\nonumber
 \\
 + \eta_{\text{cav}}(\mathbf{r},t) + \eta_{\text{ex}}(\mathbf{r},t) + \hbar\int\limits_{0}^{t}dt' \Sigma^{\text{R}}(t,t')\psi(\mathbf{r},t')
\label{sys1}
 \end{align}
In our model, the evolution of the density of the exciton reservoir
is described by the equation
\begin{align}
\frac{\partial\rho(\mathbf{r},t)}{\partial t} &{=} \frac{1}{\hbar}P_{\text{incoh}}(\mathbf{r},t) {-} \gamma_{\text{exR}}\rho(\mathbf{r},t) {-}\frac{2}{\hbar}\text{Im}\left[\psi^*(\mathbf{r},t)\eta(\mathbf{r},t)\right] - \nonumber\\
&-2\text{Im}\left[
\psi^*(\mathbf{r},t')\int\,dt'\Sigma^{\text{R}}(t,t')\psi(\mathbf{r},t')
\right],
\label{dndt}
\end{align}
where the term $P_{\text{incoh}}(\mathbf{r},t)$ describes the incoherent pumping of the reservoir, $\gamma_{\text{exR}}$ is the decay rate
of the reservoir excitons. The third term on the right-hand side describes polariton exchange between the condensate and the reservoir via the fluctuations.

In \cite{PLA} we have demonstrated that at low temperatures an exponential approximation can be used both for (\ref{retarded}) and (\ref{keldysh}):
\begin{align}
 \Sigma^{\text{R}}(t,t') &\simeq i\frac{\rho_{\text{r}}^2\alpha_{\text{c}}^2 }{\hbar^2}e^{-\gamma_{\text{eff}}(t-t')}\theta(t-t'),\\
 \Sigma^{\text{K}}(t,t') &\simeq -i\frac{\rho_{\text{r}}^2\alpha_{\text{c}}^2 }{\hbar^2}e^{-\gamma_{\text{eff}}|t-t'|}.
 \label{SK}
\end{align}
Moreover, for low temperatures, $\gamma_{\text{eff}}$ is linearly dependent on temperature \cite{PLA22}. In this case, the autocorrelation function (\ref{autocorr}) corresponds to the complex-valued stochastic Ornstein-Uhlenbeck process,
when $\tilde\eta(\mathbf{r}, t)$ is the solution of the following Langevin equation:
\begin{equation}
 \frac{d\tilde\eta(\mathbf{r}, t)}{dt} = -\gamma_{\text{eff}}\tilde\eta(\mathbf{r}, t) + \sqrt{2\gamma_{\text{eff}}}\xi(\mathbf{r}, t).
\end{equation}
Here $\xi(\mathbf{r}, t)$ is a complex white noise with unit variance,
\begin{equation}
 \braket{\xi^*(\mathbf{r}, t)\xi(\mathbf{r'}, t')} = \delta(t-t')\delta(\mathbf{r}-\mathbf{r'}).
\end{equation}
The exponential form of the memory kernel of the evolution equation (\ref{sys1}) makes if possible to transform it into an equivalent Markov equation by introducing an auxiliary wave function

\begin{equation}
 \phi(\mathbf{r},t) = \psi_0(\mathbf{r})e^{-\gamma_{\text{eff}}t}
 +\int\limits_{0}^t\,dt' e^{-\gamma_{\text{eff}}(t-t')}\psi(\mathbf{r},t'),
 \label{fictious}
\end{equation}

where $\psi_0(\mathbf{r}) = \psi(\mathbf{r},t=0)$. This technique is known as Markov embedding \cite{DeVega,Xiantao}. The equations (\ref{sys1}) and (\ref{dndt}) are now expressed as:
\begin{widetext}
\begin{align}
 i\hbar\frac{\partial\psi(\mathbf{r},t)}{\partial t} &=
 \hat H_0\psi(\mathbf{r},t) - \frac{i\hbar \gamma_{\text{cav}}}{2}\psi(\mathbf{r},t) + P_{\mathrm{coh}}(\mathbf{r},t)+ \eta(\mathbf{r},t) + 
 i\frac{\alpha_{\text{c}}^2\rho_{\text{r}}^2(\mathbf{r},t)}{\hbar\gamma_{\text{eff}}}
 \left[\phi(\mathbf{r},t) - \psi_0(\mathbf{r})e^{-\gamma_{\text{eff}}t}\right],\label{sys2}\\
\frac{\partial\rho_{text{r}}(\mathbf{r},t)}{\partial t} &= \frac{1}{\hbar}P_{{\text{incoh}}}(\mathbf{r}) - \gamma_{\text{exR}}\rho_{\text{r}}(\mathbf{r},t)-2\frac{\alpha_{\text{c}}^2\rho_{\text{r}}^2(\mathbf{r},t)}{\hbar^2}\text{Re}\left\{
\psi^*(\mathbf{r},t)\left[\phi(\mathbf{r},t) - \psi_0(\mathbf{r})e^{-\gamma_{\text{eff}}t}\right]
\right\}-\frac{2}{\hbar}\text{Im}\left\{\psi^*(\mathbf{r},t)\eta(\mathbf{r},t)\right\}.
\end{align}
These equations have to be supplemented with the evolution equation
 for the auxiliary wave function $\phi$:
 \begin{equation}
 \frac{\partial\phi(\mathbf{r},t)}{\partial t} = \gamma_{\text{eff}}\left[\psi(\mathbf{r},t) - \phi(\mathbf{r},t)\right].
 \label{dmdt}
\end{equation}
\end{widetext}

\section{Numerical simulation}
\label{sec:Numer}

    \begin{figure*}[!ht]
    \centering
    \includegraphics[width=.35\textwidth]{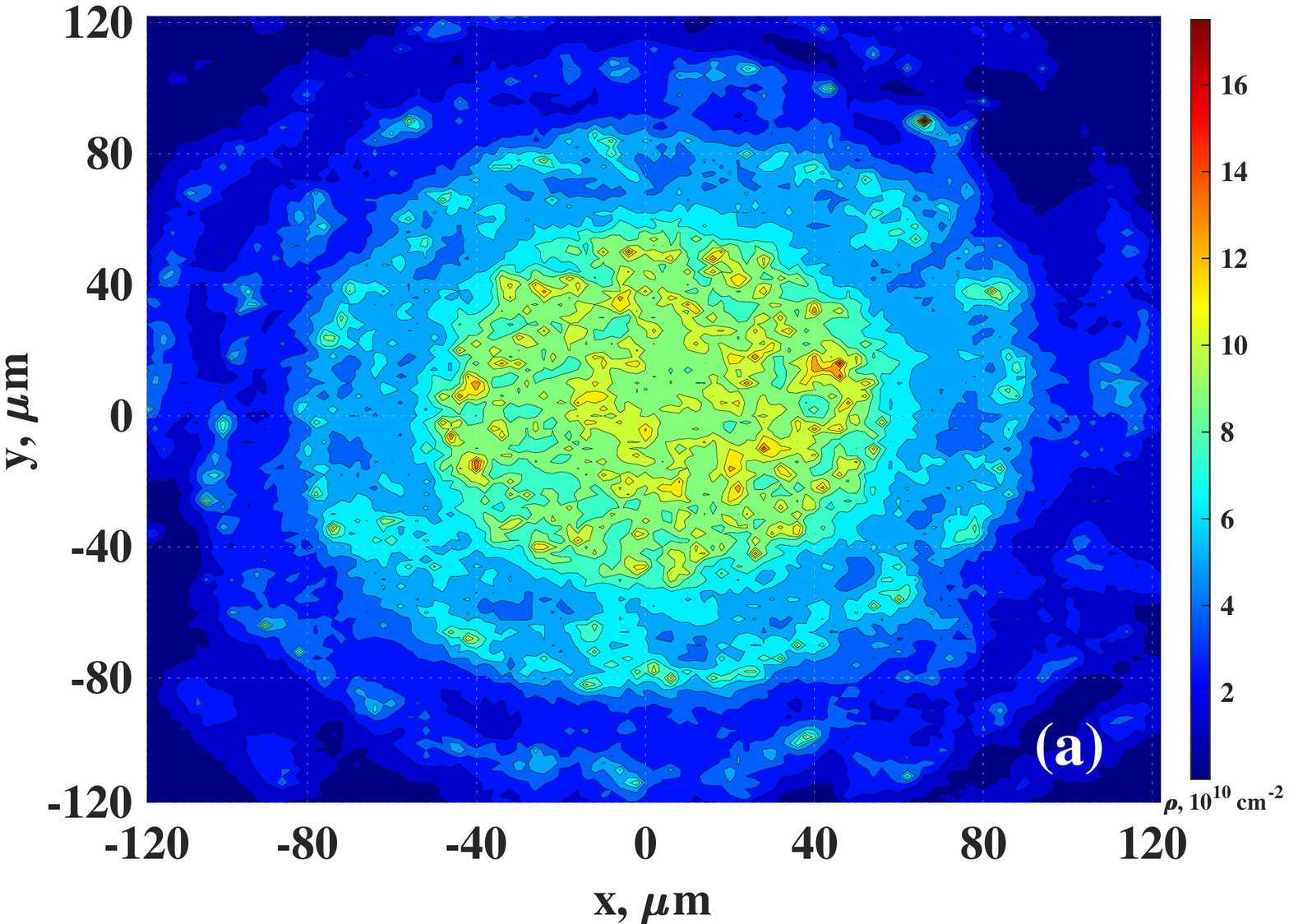}  
    \includegraphics[width=.35\textwidth]{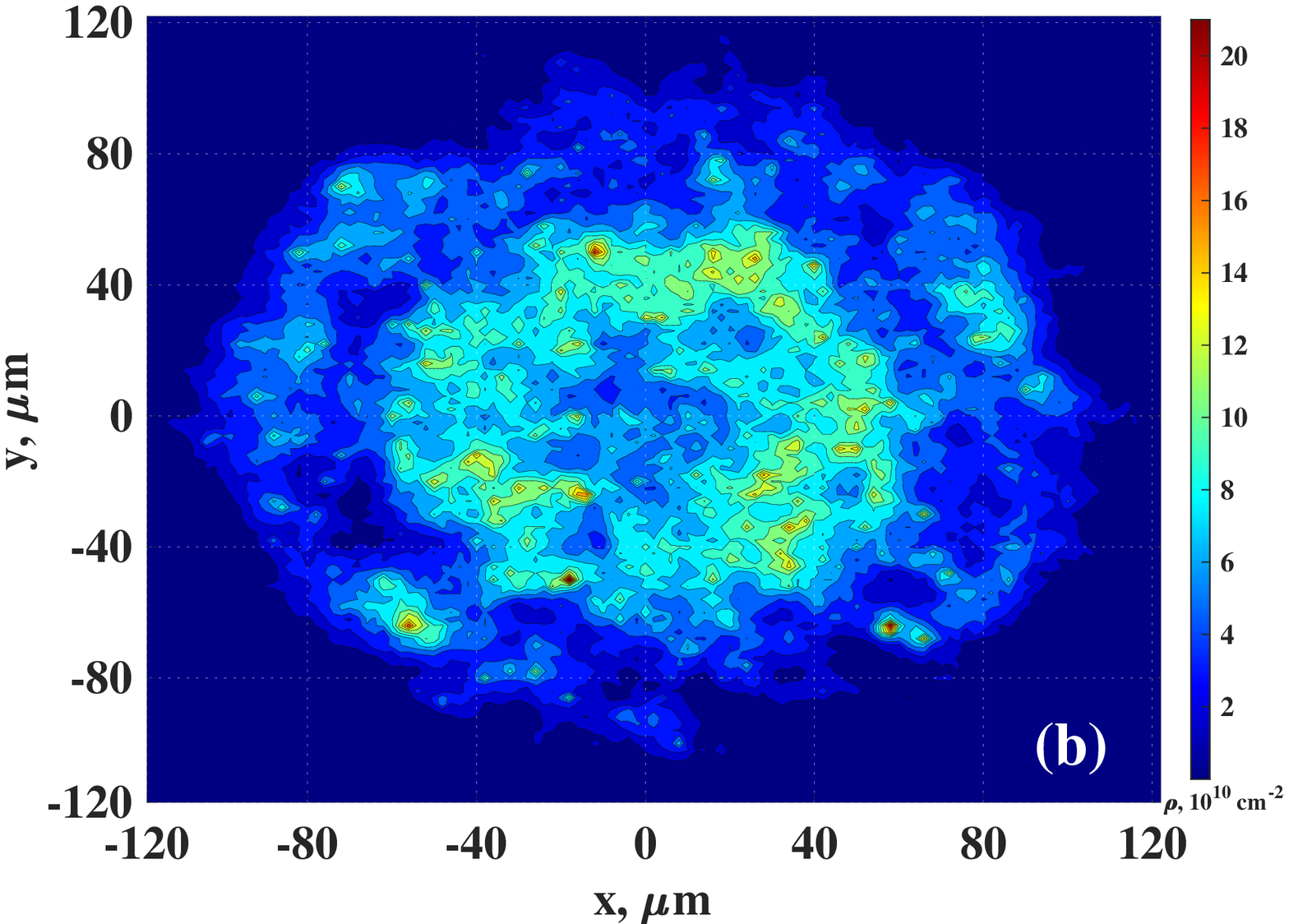}\\ 
    \includegraphics[width=.35\textwidth]{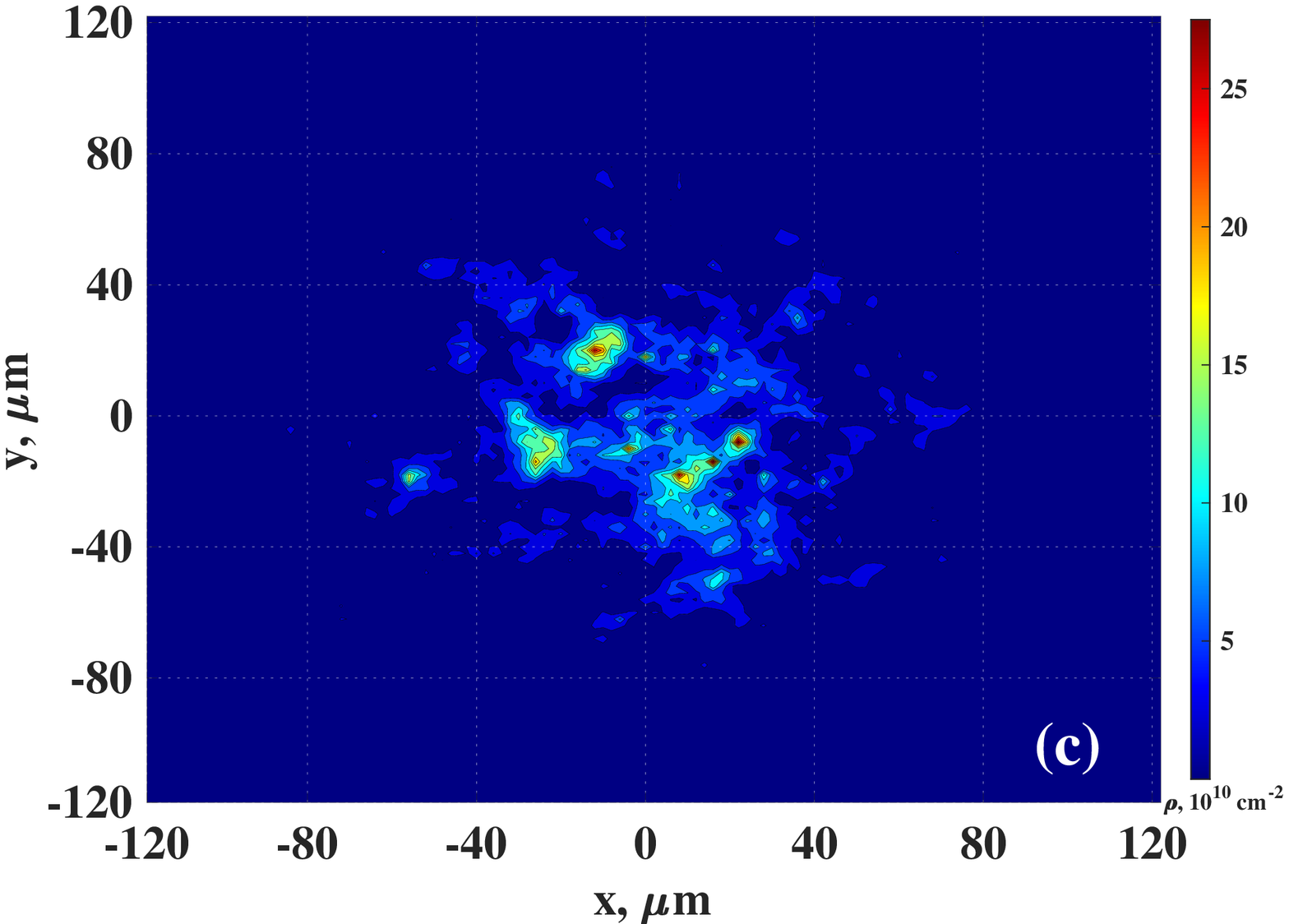}  
    \includegraphics[width=.35\textwidth]{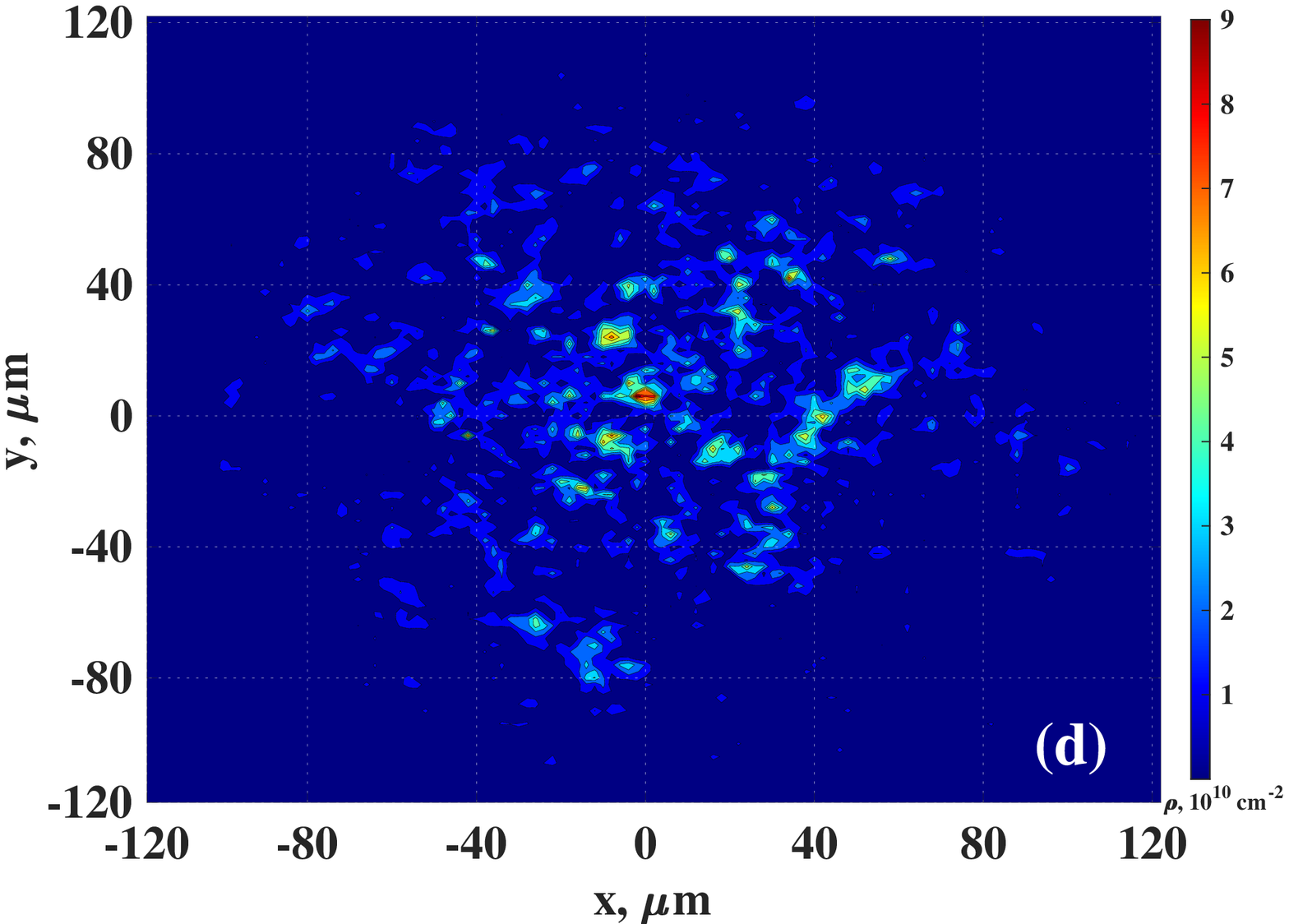} 
    \caption{Spatial distribution of condensate density at time $t=130$ ps. Temperature values: (a) 5 K, 
    (b) 20 K, (c) 35 K, (d) 50 K.
    All data presented correspond to a single realization of the fluctuation field.}
    \label{Snap_rho}
    \end{figure*}
    
In the present paper, the system of equations (\ref{sys2})--(\ref{dmdt}) is used to study condensate formation in a polariton gas exposed to incoherent
pumping, i.e., in the absence of direct coherent photon pumping into the microcavity, $P_{\text{coh}}=0$.
We consider the case of constant pumping:
\begin{equation}
 P_{\text{incoh}}(\mathbf{r}) = \gamma_{\text{exR}}\rho_0 w(\mathbf{r}),
\end{equation}
with the function $w(\mathbf{r})$ defining the spot profile
\begin{equation}
 w(\mathbf{r}) = \exp\left[-\left(\frac{\mathbf{r} - \mathbf{r_{\text{c}}}}{\sigma_{\text{{r}}}}\right)^2
 \right].
\end{equation}
We place the pump spot center at the origin, $\mathbf{r_{\text{c}}} = 0$.
The value $\rho_0$ represents the maximum reservoir density in the equilibrium state.
We use $\rho_0 = 0.5\times 10^{12}$ $cm^{-2}$ for numerical simulation.
The parameter $\sigma_{\text{r}}=20$ $\mu m$ sets the half-width of the pump spot.
The values of $\gamma_{\text{cav}}$, $\gamma_{\text{ex}}$ and $\gamma_{\text{exR}}$ and are determined by the corresponding lifetimes:
$\tau_{\text{cav}}=1/\gamma_{\text{cav}}=3.8$~ps, 
 $\tau_{\text{ex}}=1/\gamma_{\text{ex}}=1$~ps, 
 $\tau_{\text{exR}}=1/\gamma_{\text{exR}}=10$~ps.
The interexciton interaction constant $\alpha_{\text{c}}$ is set to $6\cdot 10^{-14}$~eV$\cdot$cm$^{2}$, the size of the grid cell $\Delta x=\Delta y=0.5$ $\mu m$.
We consider the case of zero initial conditions
\begin{equation}
 \psi(t=0) = \phi(t=0) = \rho_{\text{r}}(t=0) = 0,
\end{equation}
i.e., photon and exciton fluctuations play the role of the "seeds" for condensate nucleation.
In this case the condensate formation develops a process of self-organization accompanied by gradual increasing of condensate healing length. It is reasonable to assume that the dynamic memory in the Gross-Pitaevskii equation should contribute to this process. Since memory time is determined by the reservoir temperature, we expect significant influence of the latter on the dynamics of the system.
    \begin{figure*}[!ht]
    \centering
    \includegraphics[width=.35\linewidth]{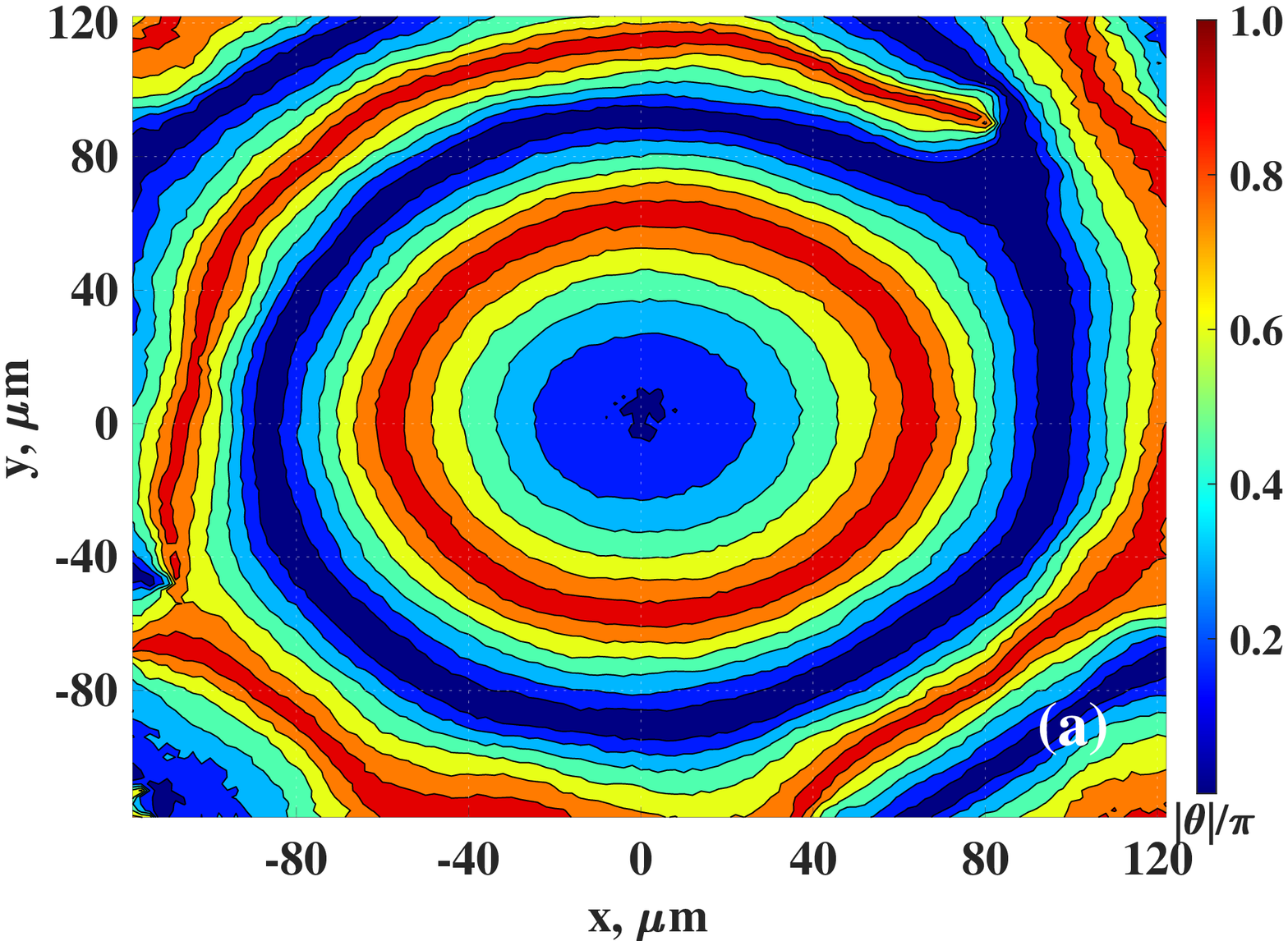}\hspace{0.1cm}  
    \includegraphics[width=.35\linewidth]{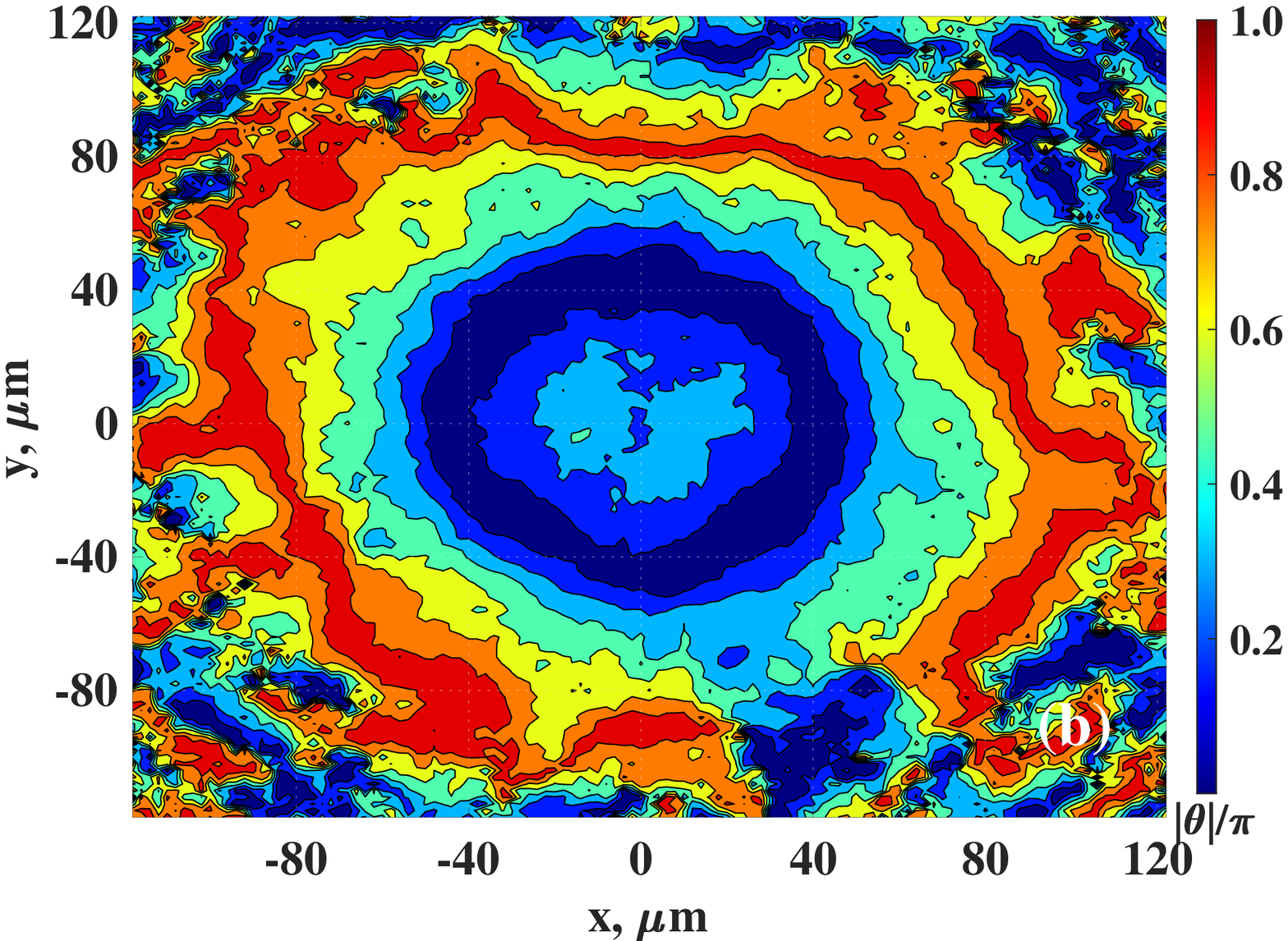}
    \vspace{1 mm}
    \includegraphics[width=.35\linewidth]{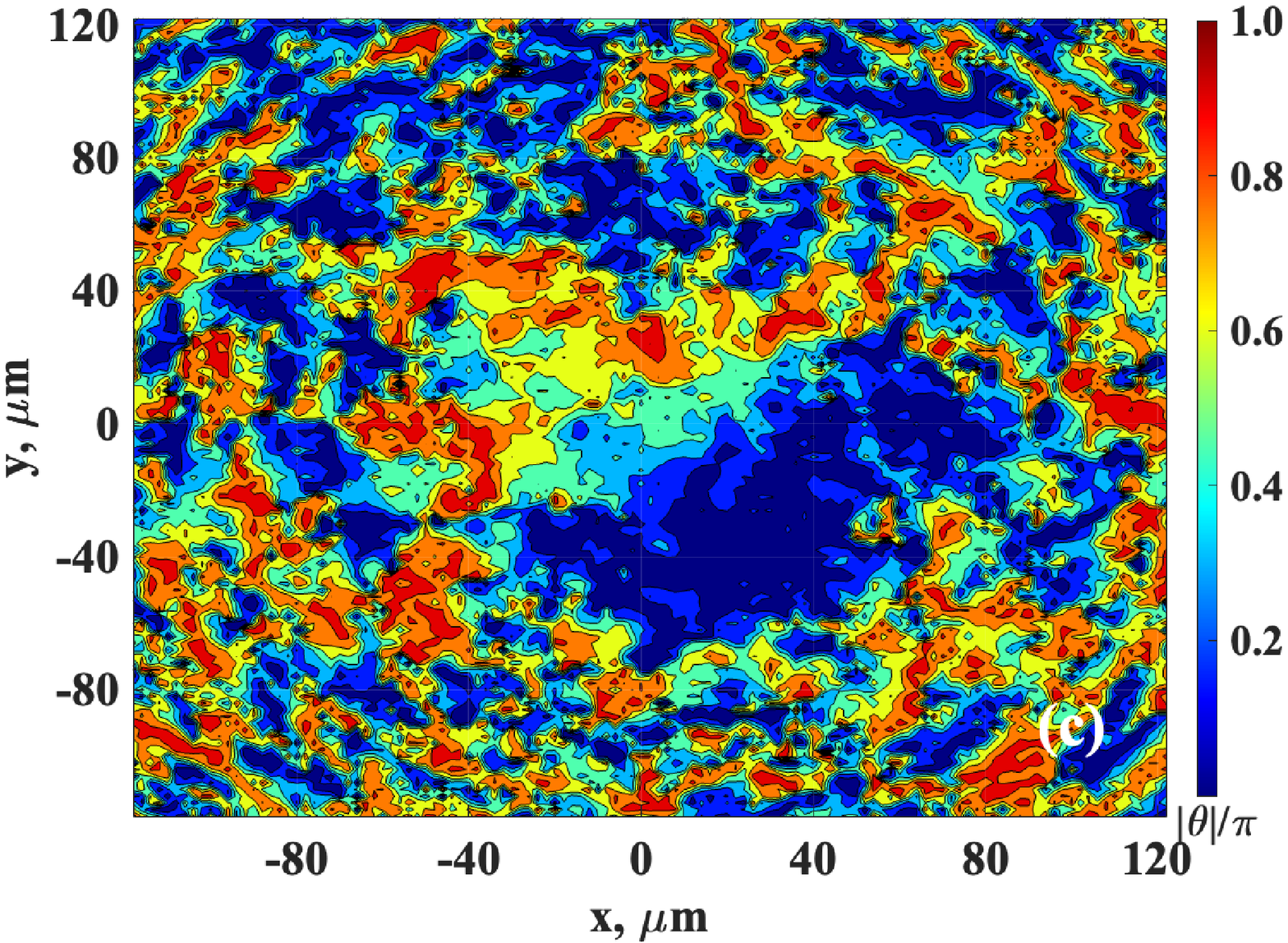}\hspace{0.1cm}
    \includegraphics[width=.35\linewidth]{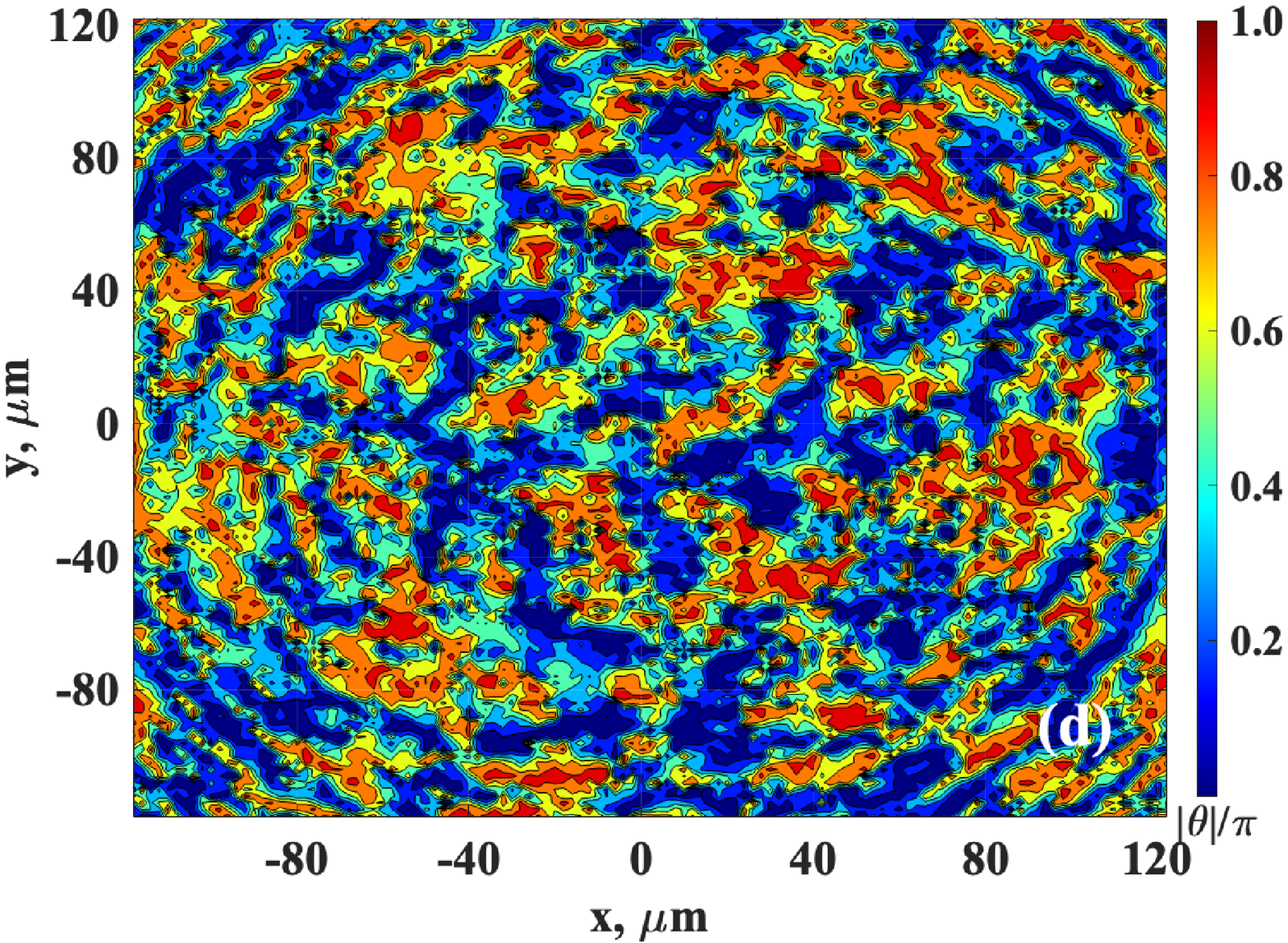}
    \caption{Spatial distribution of the condensate phase absolute value at time $t=130$ ps. The temperature values are: (a) 5 K, 
    (b) 20 K, (c) 35 K, (d) 50 K. All data presented correspond to a single realization of the fluctuation field.}
    \label{Snap_phase}
    \end{figure*}
Results of numerical simulation confirm the latter assumption.
Figure ~\ref{Snap_rho} shows condensate density distributions for different temperatures and individual realisations of $\eta_{\text{cav}}(\mathbf{r},t)$ and $\eta_{\text{ex}}(\mathbf{r},t)$. All cases correspond to $t=130$~ps. By this time, all transients have finished, and the system reaches a quasi-equilibrium state.

As follows from the data presented in Fig.~\ref{Snap_rho}, the structure of this state is qualitatively temperature dependent. At $5$ and $20$ K, the condensate occupies a vast region, with the spatial density distribution roughly repeating the intensity distribution for the incoherent pump beam. At $35$ and $50$ K, however, we observe the condensate as individual spots of irregular shape. This indicates that dynamic memory, the duration of which increases with decreasing temperature, plays a significant role in establishing the spatial correlations of the condensate.  The phenomenon of condensate fragmentation with increasing temperature was previously discussed in \cite{PLA22}.

Destruction of spatial correlations of the condensate with increasing temperature is  more clearly demonstrated in Fig. \ref{Snap_phase}, which depicts spatial condensate phase distributions for the same realizations as in Fig.~\ref{Snap_rho}.
At $T=5$~K, the lines of constant phase take the form of concentric rings, somewhat distorted due to the presence at the periphery of several singular points corresponding to the vortex cores. Such a phase configuration indicates the presence of matter waves being emitted from the pump region with a fairly regular wave front shape. At $T=20$~K the wave front undergoes significant spatial distortions due to the influence of spatial condensate fluctuations. With a further increase in temperature, the contribution of fluctuations increases, which leads eventually to the destruction of the ring phase configuration. At $T=35$~K traces of phase coherence are preserved only for individual condensate fragments, and at $T=50$~K they almost completely disappear.

The behavior of individual statistical realizations sheds light on the qualitative dependence of the condensate dynamics on the temperature, though statistical modeling is yet necessary.

As one of the indicators of the condensate state, let us consider  condensate density defined as
\begin{equation}
 \bar\rho =\frac{1}{\pi r_{\text{c}}^2} \int d\mathbf{r} f(\mathbf{r})|\psi(\mathbf{r})|^2\,
\end{equation}
where the function $f(\mathbf{r})$ extracts the regions of the pump spot with the highest density,
\begin{equation}
    f(\mathbf{r}) = \Biggl\{\Biggr.\begin{aligned}
    1,\quad |\mathbf{r}-\mathbf{r}_{\text{c}}|\le \sigma, \\
    0, \quad |\mathbf{r}-\mathbf{r}_{\text{c}}| > \sigma.
    \end{aligned}
\end{equation}

    \begin{figure}[!ht]
    \centering
    \includegraphics[width=.98\linewidth]{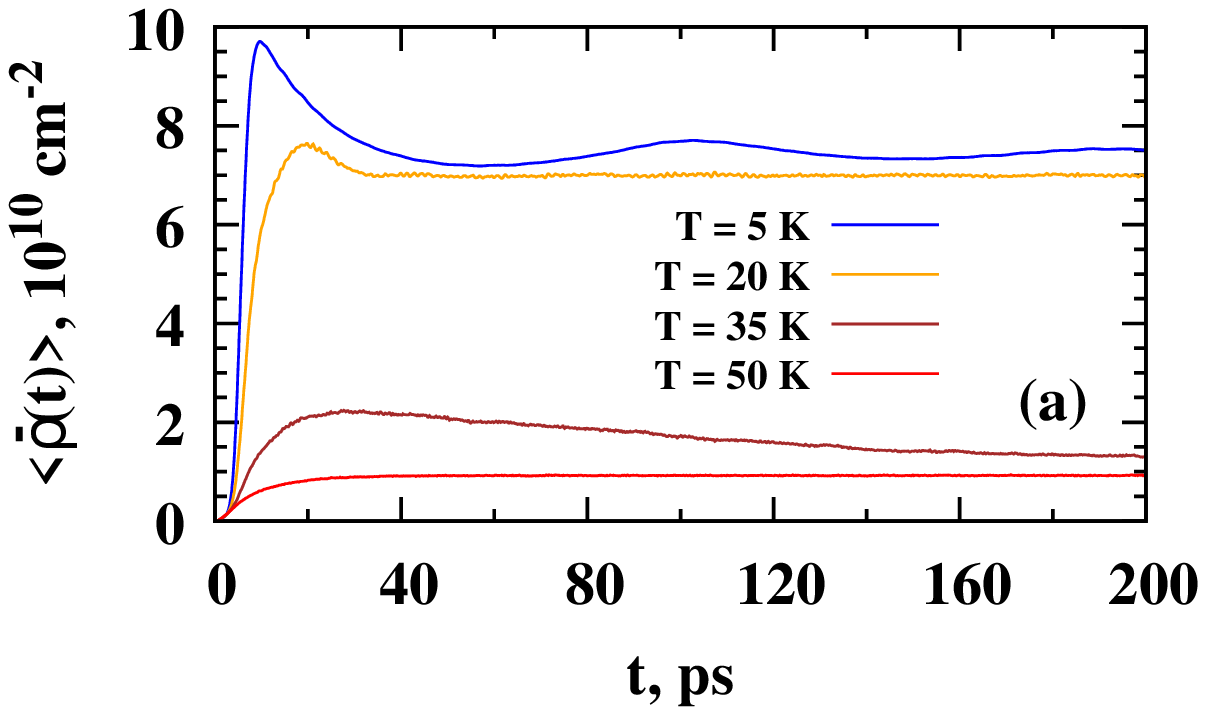}  
    \includegraphics[width=.98\linewidth]{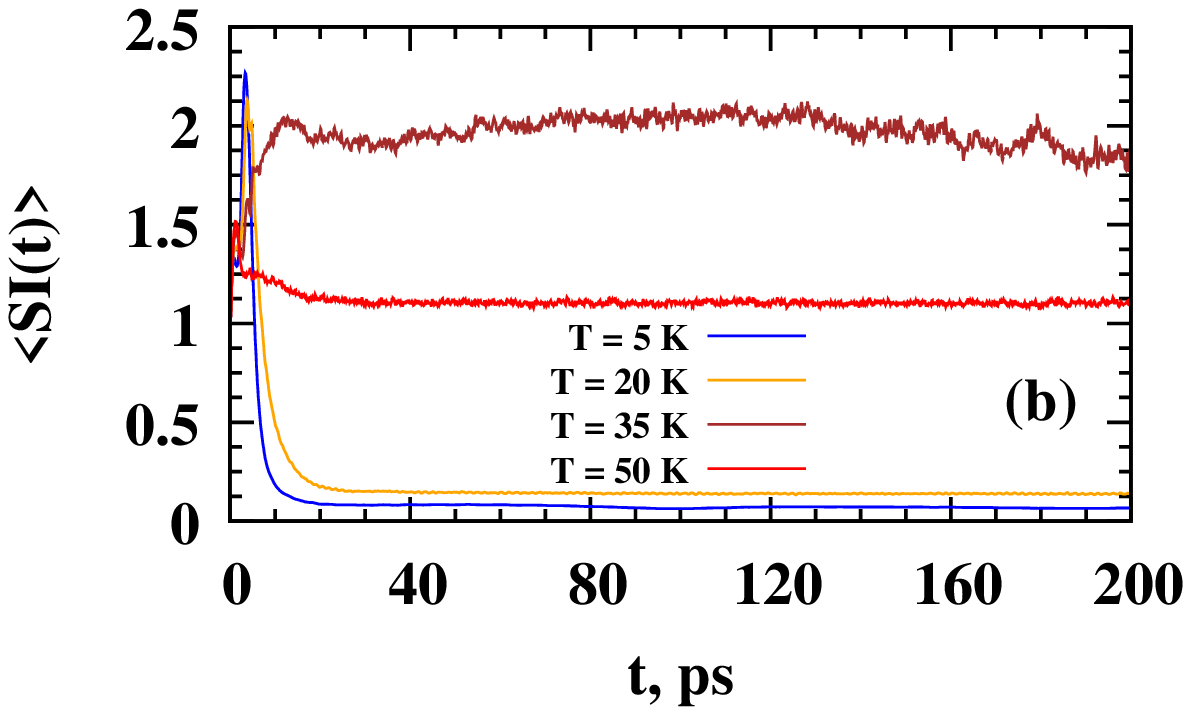} 
    \caption{Time dependencies of ensemble-averaged density $\bar\rho$ (panel (a))
     and the scintillation index (panel (b)) for different temperature values.}
    \label{Rho_SI}
    \end{figure}

We use also the scintillation index as a characteristics of the spatial condensate distribution
\begin{equation}
 \text{SI} = \frac{\braket{\bar\rho^2}}{\braket{\bar\rho}^2} - 1.
\end{equation}
The time dependence of $\bar\rho$ and scintillation index are presented in
Fig.~\ref{Rho_SI}. First of all, we note that the low-temperature phase-ordered states (5 K and 20 K) are described by significantly high densities and very weak spatial fluctuations. The sharp decrease in density that occurs with increasing temperature is accompanied by enhancement of fluctuations: we observe a pattern corresponding to the Berezinskii–Kosterlitz–Thouless(BKT) phase transition. At temperature $T=35$~K the condensate exists in the form of separate spots of high density on a turbulent background. This leads to high values of scintillation index, $\text{SI}\simeq 2$.
At $T=50$~K the contribution of such spots significantly decreases. As a consequence, the scintillation index approaches unity, which corresponds to the statistical saturation regime \cite{Dashen}. Also note that even after the condensate density stabilizes, it continues to exhibit slow oscillations. This may be due to the so-called relaxation oscillations \cite{PLA,DeGiorgy,Opala,Tian}.

The transition of the condensate to a quasi-equilibrium regime with constant or slowly varying density corresponds to the establishment of a balance between pumping from the reservoir and decay. The presence of such a balance may indicate the emergence of a state with PT (parity-time) symmetry \cite{Chestnov}. In this case the energy spectrum becomes real despite the non-Hermiticity of exciton-polaritons. The appearance of PT symmetry in a system with memory has been considered earlier in \cite{Cochran}.
    \begin{figure}[H]
    \centering
    \includegraphics[width=.75\linewidth]{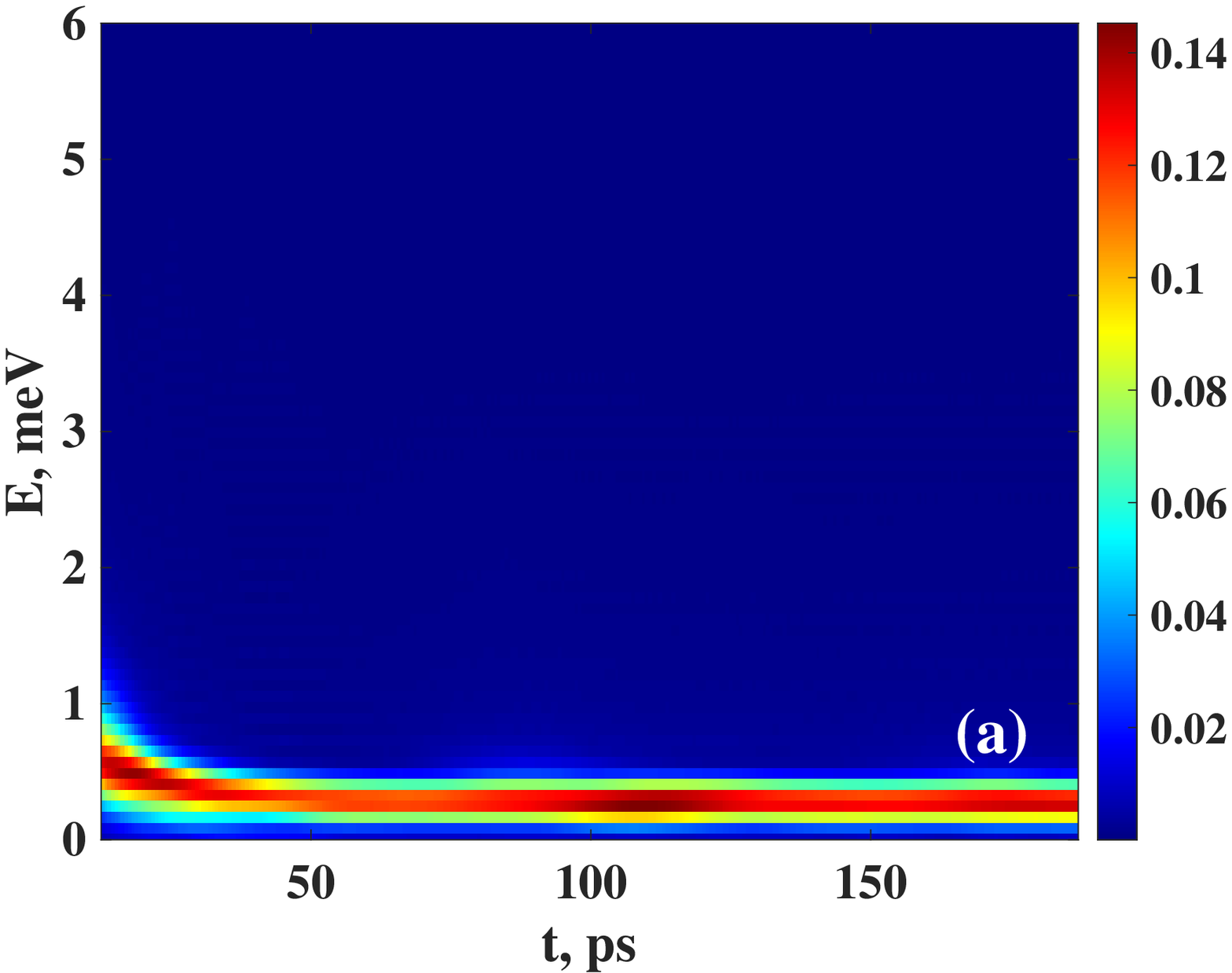}\\\hspace{0.1cm}  
    \includegraphics[width=.75\linewidth]{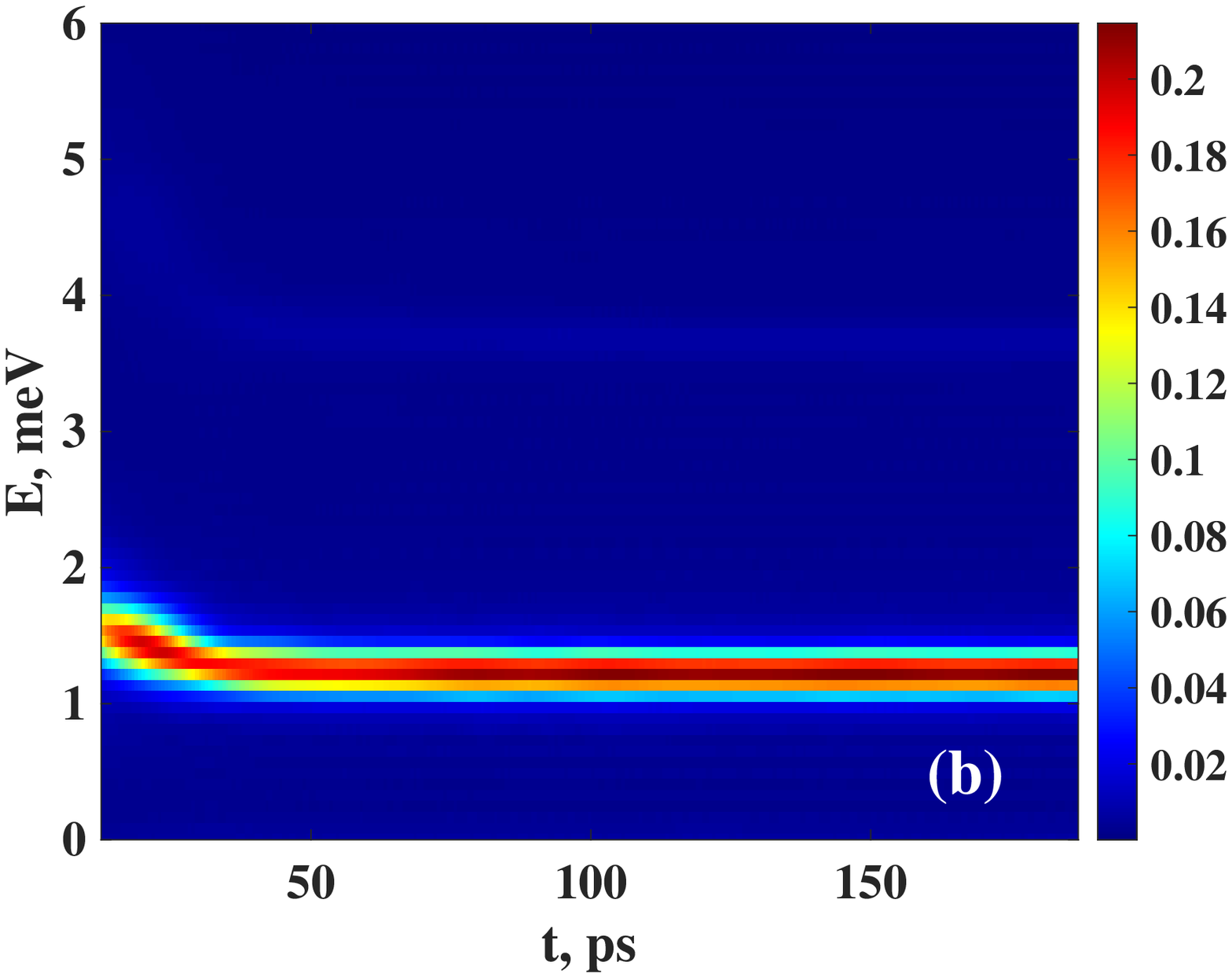}\\
    \vspace{1 mm}
    \includegraphics[width=.75\linewidth]{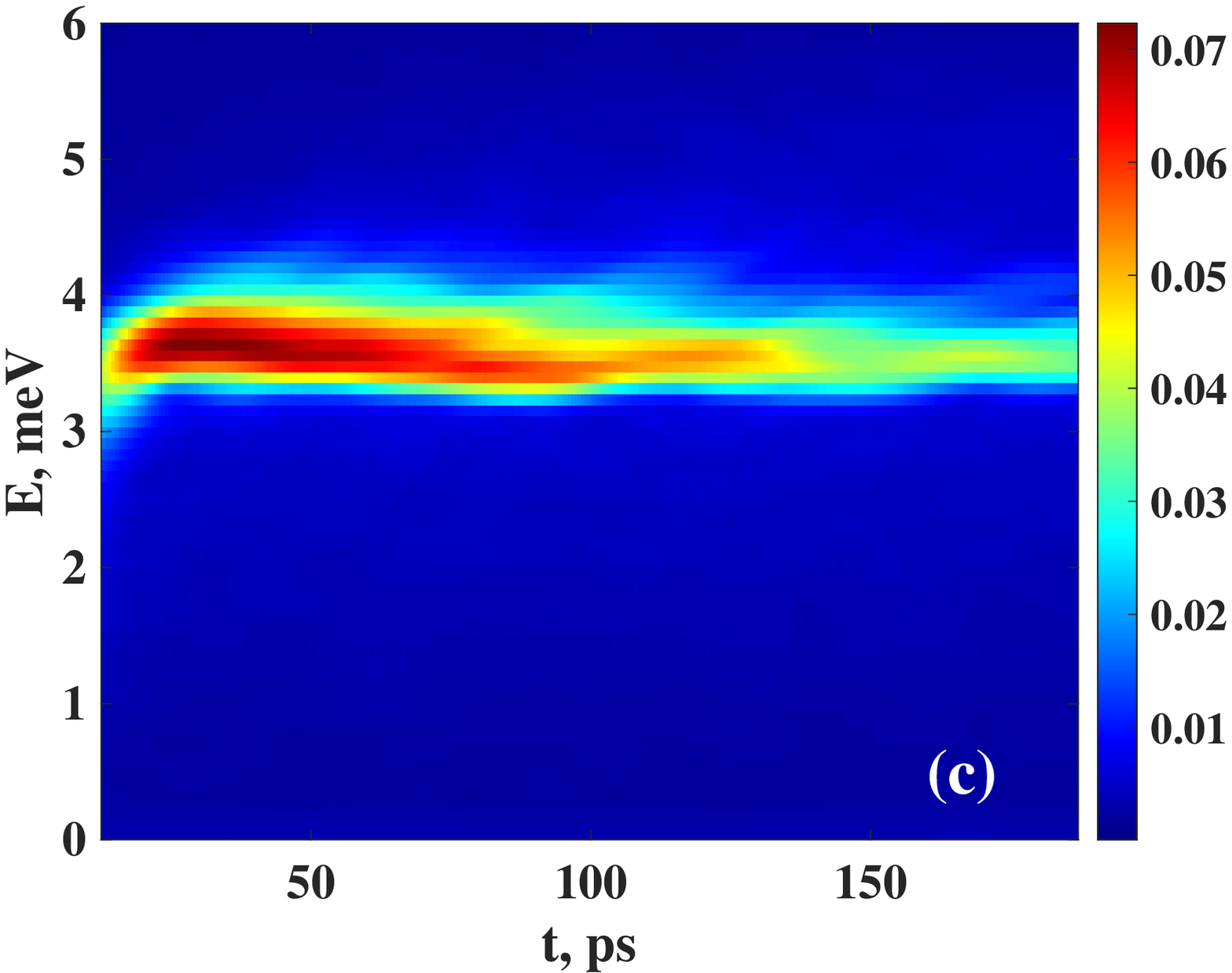}\\\hspace{0.1cm}  
    \includegraphics[width=.75\linewidth]{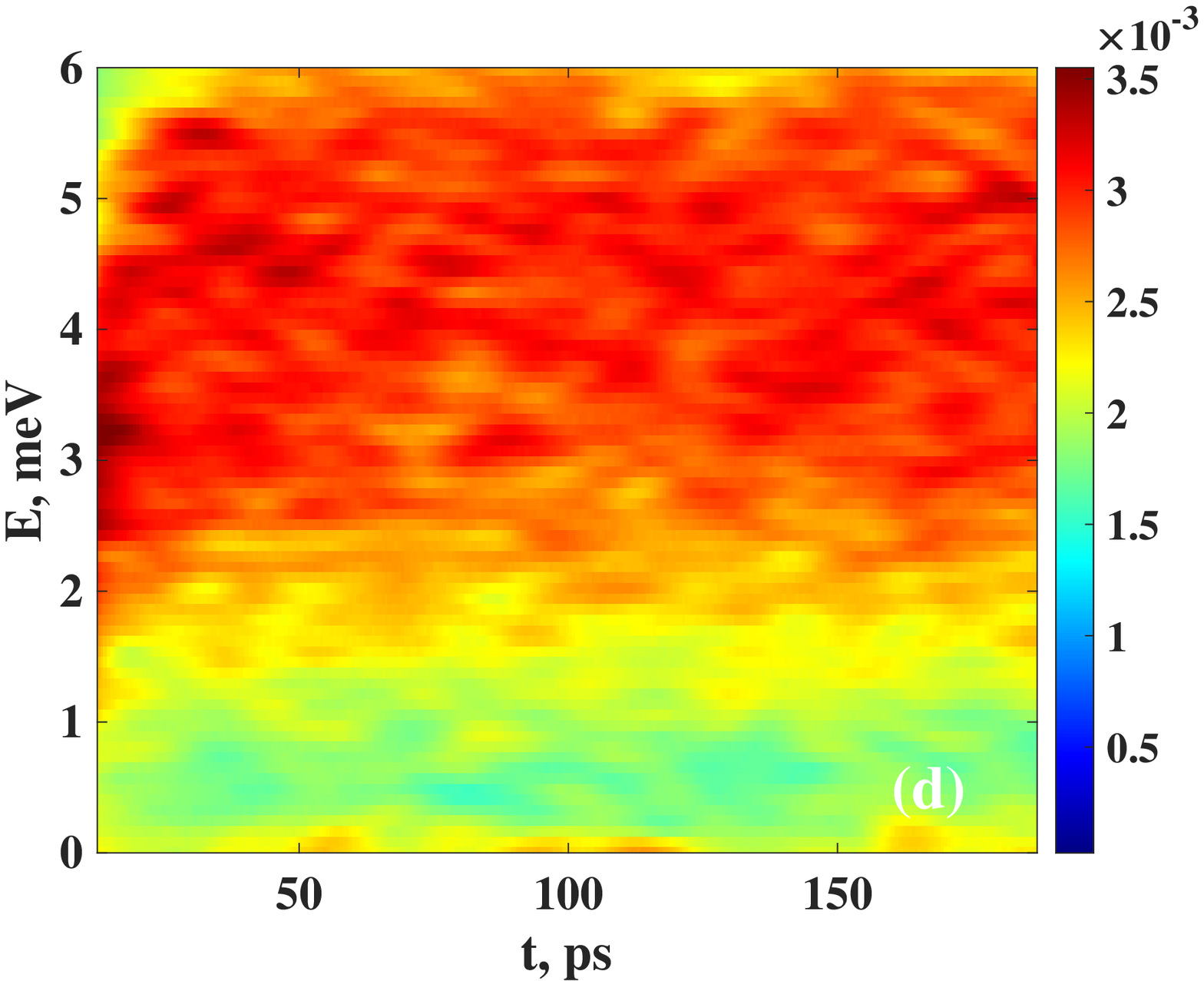} 
    \caption{Optical radiation spectrum of the condensate obtained using the Gabor transform. Temperature values: (a) 5 K, (b) 20 K, (c) 35 K, (d) 50 K. }
    \label{Gabor}
    \end{figure}

In order to trace changes in the spectrum during condensate formation, we studied the optical radiation of the condensate. To simplify the analysis, we get rid of the spatial dependence of this signal by using a weighted integration, with the weight function being given by the pumping profile:
    \begin{equation}
        \Psi_{\text{eff}}(t) = 
        \frac{\int d\mathbf{r} w(\mathbf{r})\psi(\mathbf{r},t)}{\int d\mathbf{r} w(\mathbf{r})}.
    \end{equation}
To obtain the spectrum of the signal, we use the Gabor transform, which is a variation of the Fourier window transform with the Gaussian window:
    \begin{equation*}
        g_{W}(\nu,t) = \int_{-\infty}^{\infty}dt^{\prime} \Psi_{\text{eff}}(t^{\prime})\chi_{\text{W}}(\nu,t^{\prime}-t),
    \end{equation*}
where 
%
    \begin{equation}
        \chi_{\text{W}}(\nu,t) = R_{\text{W}}(t)\exp{(2\pi i\nu t)}
        \label{chi}
    \end{equation}
and
    \begin{equation}
        R_{\text{W}}(t) = \frac{1}{\sqrt{W\sqrt{2\pi}}}\exp{\left(-\frac{t^2}{4W^2}\right)}.
    \end{equation}

The window is centered at $t = t^{\prime}$, and $W$ sets its size. By moving the center of the window along the time axis, one may trace the time dependence of the frequency spectrum of $\Psi_{\text{eff}}(t)$. Fig.~\ref{Gabor} shows the spectral density of the signal averaged over 100 realizations. One may observe that at temperatures 5 K and 20 K the condensate is concentrated in low-energy states within a narrow spectral band. The mean frequency of this band corresponds to the equilibrium chemical potential. Its increase with increasing temperature is associated with enhancement of the condensate inhomogeneity and, as a consequence, higher contribution of kinetic energy. At 35 K, the condensate spectrum is still given by a fairly narrow band, but the shape and intensity of this band changes with time. Comparing the evolution of the spectrum with the data presented in Fig.~\ref{Rho_SI}(a), one may conclude that the decay is not completely compensated by pumping. This may be due to the metastability of condensate spots arising at 35 K. The spectrum at 50 K is qualitatively different from the previous cases: instead of excitation of a narrow spectral band, we observe a broadband spectrum with irregular variability with time. Such behavior is consistent with the pattern of disordered turbulent condensate dynamics.    

\section{Conclusion}
\label{sec:Concl} 

Let's summarize the main results of the paper. We have presented a non-Markovian stochastic Gross-Pitaevskii model for the exciton-polariton Bose-Einstein condensation. We used it to study the influence of the temperature on the condensate formation process in the absence of coherent pumping. The results obtained reveal a the complex structure of the transition from the coherent to the disordered phase. In particular, one may note the appearance of dense condensate spots on the turbulent background, detected at 35 K. Due to the presence of these spots, the spectrum of the optical signal emitted by the condensate is a narrow-band one. Further work in this field will be devoted to the improvement of the model used to describe the condensate and the reservoir.

\begin{acknowledgments}
This work was supported by the Basic Research Program of the Institute of Experimental Problems of the Far East Branch of the Russian Academy of Sciences, project no. 121021700341-2 and by the foundation for the advancement of theoretical physics and mathematics “Basis”. The work of Yu. E. Lozovik and N.A. Asriyan is supported by RFBR projects 20-02-00410 and 21-52-12038.
\end{acknowledgments}

\end{document}